\documentclass[twocolumn]{aastex631}
\usepackage{graphicx}  
\usepackage{bm}        
\usepackage{amsfonts,amsmath,amssymb,mathrsfs}
\usepackage{hyperref}
\usepackage{xcolor}
\makeatletter
\let\saved@includegraphics\includegraphics
\AtBeginDocument{\let\includegraphics\saved@includegraphics}
\makeatother

\usepackage{color}
\usepackage{times}
\hypersetup{
  colorlinks=true,        
  linkcolor=blue,         
  citecolor=magenta,      
}

\newcommand\aastex{AAS\TeX}
\newcommand\latex{La\TeX}

\shorttitle{The accurate mass distribution of M87}
\shortauthors{De Laurentis \& Salucci}

\graphicspath{{./}{figures/}}

\begin{document}

\title{The accurate mass distribution of M87, the Giant Galaxy with  imaged shadow of its supermassive black hole, as a portal to new Physics\footnote{Released on March, 1st, 2021}}

\correspondingauthor{Mariafelicia De Laurentis}
\email{mariafelicia.delaurentis@unina.it, salucci@sissa.it}

\author[0000-0002-
9945-682X]{Mariafelicia De Laurentis}
\affiliation{Dipartimento di Fisica “E. Pancini,”\\
Universitá di Napoli “Federico II”\\
Complesso Universitario di Monte S. Angelo, Edificio G, Via Cinthia,\\ I-80126 Napoli, Italy}
\affiliation{INFN Sez. di Napoli,\\ Complesso Universitario di Monte S. Angelo, Edificio G, Via Cinthia,\\ I-80126 Napoli, Italy}
\author[0000-0002-5476-2954]{Paolo Salucci}
\affiliation{SISSA International School for Advanced Studies,\\ Via Bonomea 265,\\ 34136, Trieste, Italy}
\affiliation{INFN, Sezione di Trieste,\\ Via Valerio 2, 34127,\\ Trieste, Italy}

\begin{abstract}

The very careful Event Horizon Telescope estimate of the mass of the supermassive black hole at the center of the Giant CD galaxy M87, allied with recent high quality photometric and spectroscopic measurements, yields a proper dark/luminous mass decomposition from the galaxy center to its virial radius. 
That provides us with decisive information on crucial cosmological and astrophysical issues. 
The dark and the standard matter distributions in a wide first time detected galaxy region under the supermassive black hole gravitational control. 
The well known supermassive black hole mass vs stellar dispersion velocity relationship at the highest galaxy masses implies an exotic growth of the former. This may be the first case in which one can argue that the supermassive black hole mass growth was also contributed by the Dark Matter component. A huge dark matter halo core in a galaxy with inefficient baryonic feedback is present and consequently constrains the nature of the dark halo particles.  
The unexplained entanglement between dark/luminous structural properties, already  emerged in disk systems, also appears.
\end{abstract}

\keywords{nnnn --- ooooo --- History of astronomy(1868) --- Interdisciplinary astronomy(804)}

\section{Introduction} \label{1}
The distribution of Dark Matter (DM) in galaxies is extremely relevant for Cosmology and Particle Physics. 
Let us sketch the state of the art: the well-known ${\rm \Lambda}$CDM scenario predicts, employing cosmological N-Body simulations, that the DM halo density, in any virialized object, follows the Navarro-Frenk-White (NFW) profile \citep{Navarro:1995iw}, characterized by a central cusp: $\rho(r)\propto  r^{-1} (r+r_s)^{-2}$ with $r_s$ a length scale depending on the value of the halo mass.  
However, the individual and coadded kinematics of Spirals, Low Surface Brightness (LSB) galaxies and Dwarf Irregulars clearly show that the main baryonic component, a stellar disk of surface density\footnote{The size of the stellar disk is defined as $R_{opt}\equiv 3.2 \ R_D$} \citep{Freeman:1970mx}: 
\begin{equation}
\label{Sigma}
\Sigma_{\star}(r)= \frac{M_D}{2 \pi R_D^2} e^{-r/R_D}\,,
\end{equation}
is embedded in a dark halo with  a {\it cored} density distribution\cite{dB10,Svat,Karukes_2016,Salucci_2019}: $\rho(r) \propto (r_0+ r)^{-1} (r_0^2+ r^2)^{-1}$.  
The above discrepancy between the empirical profile and the $\Lambda$CDM N-Body outcome simulation is particularly strong in dwarfs and low luminosity disc systems \cite{Karukes_2016}. Remarkably, after a proper circular-velocity decomposition into its dark and luminous components, all disk systems, of stellar masses $ 5\times 10^7 \ M_\odot \leq M_\star  \leq 3 \times  10^{11} M_\odot $, show a cored dark halo density profile (see also  \citep{deMartino:2020gfi}). This is also shown by coadded studies of rotations curves \citep{Salucci_2007a, Salucci_2007b, Dehghani2020cvl} (see also \href{https://www.youtube.com/playlist?list=PLe9EzUPIhHubf_m-F7eK10t7A4MIg4fIu}{\color{blue}Di Paolo 2021}). At high redshifts the situation is open although some evidences for dark matter cores are appearing \cite{Salucci:2020nlp}. However, it is worth noticing that supernovae explosions in the stellar disks may originate a "baryonic feedback" capable of fabricating the observed DM cores from the original cusps \cite{dc14}. 
In these systems,  however, another "anomaly" is still unexplained: the structural parameters of the mass distribution, i.e. $\rho_0, r_0$, and the disk mass and length scale  $ M_D, R_D$, are surprisingly very well correlated among themselves \citep{Salucci_2019}.

The above discussion introduces the primary goal of this work: to derive the mass distribution of the luminous and dark components of M$87$. 
This massive cD elliptical galaxy, located at the center of Virgo Cluster, is the biggest one in the Universe within a radius of $0.5$ Gpc. In detail:  its stellar spheroid mass is $20$ times bigger than that of the disk of our Galaxy, while its dark halo mass is $100$ times more massive.  
We will then investigate the galaxy structural properties in an extreme object.  

Moreover, as any spheroidal galaxy, M$87$  has, at its center, a very supermassive black hole (SMBH), its mass estimates, obtained from different methods/measurements, have ranged from $(3.5 \pm 0.8) \times 10^9 M_{\odot}$ to $7.22^{+0.34}_{-0.40}\times 10^9 M_{\odot}$ \citep{Walsh:2013uua,Oldham2016}.  
This SMBH is the first and the only one so far to be imaged (as published in April 2019 \citep{Akiyama:2019cqa,Akiyama:2019eap}). 
The image shows its shadow, surrounded by an asymmetric emission ring with a diameter of $3.36\times 10^{-3}$\ pc ($0.01$ ly). 
This result has a consequence for the present work: as a byproduct, the Event Horizon Telescope (EHT) has finely measured the SMBH mass: $(6.5\pm 0.2_{\rm stat}\pm 0.7_{\rm sys} )\times 10^9 M_{\odot}$ \citep{Akiyama:2019eap}. 
Its value and small uncertainty play an essential role in the present work results. Finally, we assume for M$87$ the EHT distance of $16.5$ Mpc \citep{Akiyama:2019cqa}.

This work is organised as follows. In Section \ref{2}, we describe the method obtained to derive the M87 mass model accurately.
In the next Section \ref{3}, we investigate the properties of the central SMBH and of  the other mass components of this giant galaxy.  
In the conclusions, Sec.\ref{4}, we summarize our findings specifying how M87 serves as a cosmic laboratory for many unsolved mysteries of the Universe.
\section{Method: Deriving the mass distribution in M87}
\label{2}
 
The giant elliptical M$87$ is home to several different dynamical populations as Planetary Nebulae, Globular Clusters and Satellite Galaxies. They can be used as multiple, independent and well extended tracers of the galaxy gravitational field (see e.g. \citep{A14,staar,Sch9, Napolitano:2014hda}). 
This wealth of data, alongside high-resolution photometry and spectroscopy, allows careful mass modelling, also because, in this object, we can assume a spherical symmetry in all the mass components. 
The mass structure of M87 has been the subject of many studies \citep{Strader:2011da,mga11,Z14}. Recently, a substantial improvement on  the determination of its  gravitational potential has been obtained by \citet{Oldham2} for a  region extending from very near to the center out to the virial radius. 

In relation with the galaxy DM halo density profile they adopted four models \citep{Oldham2}: one cusped, the well-known NFW profile \citep{1996NFW} and three cored: the LOG, the gNFW and the cgNFW ones. 
They found that the mass model with the DM cored profiles fared better in fitting the observational data than the cuspy ones \citep{1996NFW}.  
However, the above cored profiles raise doubts that they can describe the DM distribution in galaxies in a physically correct way. 
In fact, the LOG profile, introduced in the Universal Rotation Curve of Spirals \citep{Persic:1995ru}, holds only out to their optical radii, outside which kinematical and weak lensing data strongly favour the Burkert profile \citep{Salucci:1998ij,gentile2007ngc,Salucci_2019}, which declines with radius in the outermost halo regions (e.g. \citep{gentile2007ngc}), while the LOG profile flattens there.         
The gNFW profile features an NFW profile with the addition of a core: 
\begin{equation}
\rho(r) = \dfrac{\rho_0}{\left(\frac{r}{r_s}  \right)^\alpha \left( 1+\frac{r}{r_s}  \right)^{3-\alpha}}\,,\nonumber
\end{equation}

 $\alpha$ is the inner slope, and  $x =r/r_s$ with $r_s$ a length scale. 
However, this generalization of the NFW profile not only leads to multiple degeneracies \citep{klypin2001resolving} but also increases, by one, the number of halo parameters, with no physical justification. In fact, in both simulations and observations, the resulting DM density profiles have just two free parameters (a halo density and a halo length scale) and seem to not require a third. Actually, in the above one finds a tight relationship connecting the two, reducing so the number of necessary parameters to just one \citep{Salucci_2019}. Finally, at large radii, this density profile does not necessarily converge to the NFW one.
Similar arguments hold for the cgNFW profile also used in \citet{Oldham2} that, besides, has two free parameters more than the NFW one.

Remarkably, in Spirals, LSBs, dwarf Irregulars, dwarf Spheroidals and also ellipticals, the Burkert DM halo profile (not considered in \cite{Oldham2} alongside with its luminous counterparts fits all the available kinematics, including individual and coadded RCs excellently \citep{Salucci_2019}. 
Therefore, in this work, we adopt the latter profile and so we can also  compare its dark structure with that of galaxies of different Hubble types and halo mass. 
 
The very recent study by \citet{Oldham2} has derived the total mass distribution of M87 up to its virial radius at $1.3$ Mpc. The stellar photometry, available out to $210$ kpc, in \citet{Oldham1}, shows a 2D-light distribution which is very different from the Sersic profile, generally found in Ellipticals. We have an extended envelope, outside the inner stellar core, likely due to the fact that the  M87 spheroid  has been formed by a first burst of star formation, followed by a continuous infall of luminous matter lasting several Gyrs. The Nuker profile best models the surface luminosity profile of the stellar spheroid
\begin{equation}
I(R)  =I_0 \left(\frac{R}{R_b} \right)^{-\zeta} \left( 1+\left[ \frac{R}{R_b}\right]^{\alpha}\right)^{\frac{\zeta - \eta}{\alpha}}\,,
\label{Nuker}
\end{equation}
where, by following \citet{Oldham1} we have: $I_0= 3.5\times 10^9\  L_\odot$ kpc$^{-2}$, $\zeta =0.186$, $\eta=1.88$,  $r_b=1.05$ kpc and $\alpha=1.27$. Moreover, in M87: $M_B=-20.5$, and the half light radius reaches the value of  $r_e= (73.5 \pm 7)$. It is important to remark that Eq.\eqref{Nuker} provides us with an accurate stellar mass distribution also in the region $1 \ {\rm kpc} <r< 5{\rm kpc}$, neighbouring the central SMBH.   
From Eq.\eqref{Nuker} we derive the corresponding M87 "disk scale length" $R_D$ in the following way: in spirals, $R_D$ is the radius  inside which  the fraction of light is $0.2$ times the total one, so we get: $R_D^{M87}=10\pm 1$ kpc. 
This galaxy has one of the largest half-light radii within the region wide $1$Gpc$^3$ of the Universe. However, the value of  $R_D^{M87}$ is in line with that of the most massive spirals. 
One can argue that the peculiarity in the M87 stellar distribution emerges only at $r>50$ kpc, where the dark matter largely dominates the total mass distribution.

The mass profile  of the luminous component is obtained from Eq. \eqref{Nuker} by means of:
\begin{equation}
M_{\star}(r) = -\left(\frac{M_{sph}}{L_B}\right) \int_0^r  r^2 \left(\int_r^\infty  \left( \frac{d I(R) }{dR} \right) \frac{dR}{\sqrt{R^2-r^2}}\right) dr\,.
\label{lum mass}
\end{equation}
 
This profile has just one free parameter: $M_{sph}$ the spheroid mass. 

\begin{figure}[h!]
\centering
\includegraphics[scale=0.55]{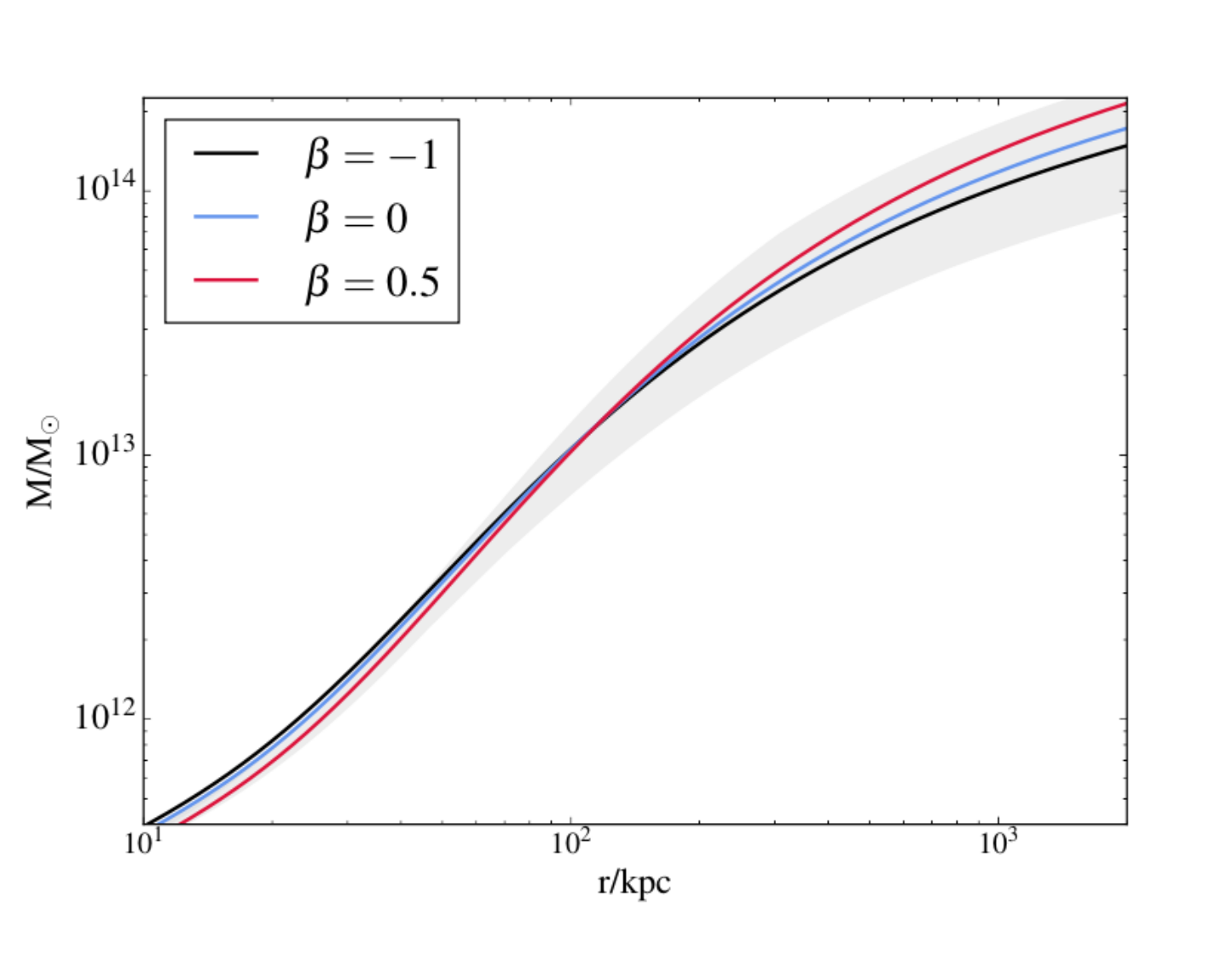}
\caption{The mass profile of the total mass of M$87$, $\beta$ is the standard anisotropy parameter (see \cite{Oldham2}).}
\label{fig:massoldham}
\end{figure}

The gravitating mass profile $M(r)$ that we use in this work was obtained from the Jeans method applied to the kinematics of different tracers of the gravitational field which assure for M87 eleven independent data points(see \cite{Oldham2}). 
They have derived the mass profile $M(r)$ for the case of spherical symmetry and also found that possible kinematical anisotropies induce only a moderate effect in the mass profile. 
In fact, by varying the anisotropy content, the corresponding mass profiles $log M(r)$ result all very similar and lie, at any radius $r$, within $0.2$ dex from the  isotropic solution \citep{Oldham2}, see Fig \ref{fig:massoldham}. 
It is worth noticing that these uncertainties in $log M(r)$ are much smaller than the range of the measured  $log M(r)$ masses that is $2.5$ dex, this opportunity is very rare for spheroidal galaxies. Therefore, in this work, we will adopt the isotropic profile $log M(r)$ of \cite{Oldham2} to represent the gravitating mass profile of M87, and we consider the effect of kinematical anisotropies by assigning a $0.2$ dex uncertainty to the log mass measurements.

\begin{figure}
\centering
\includegraphics[scale=0.4]{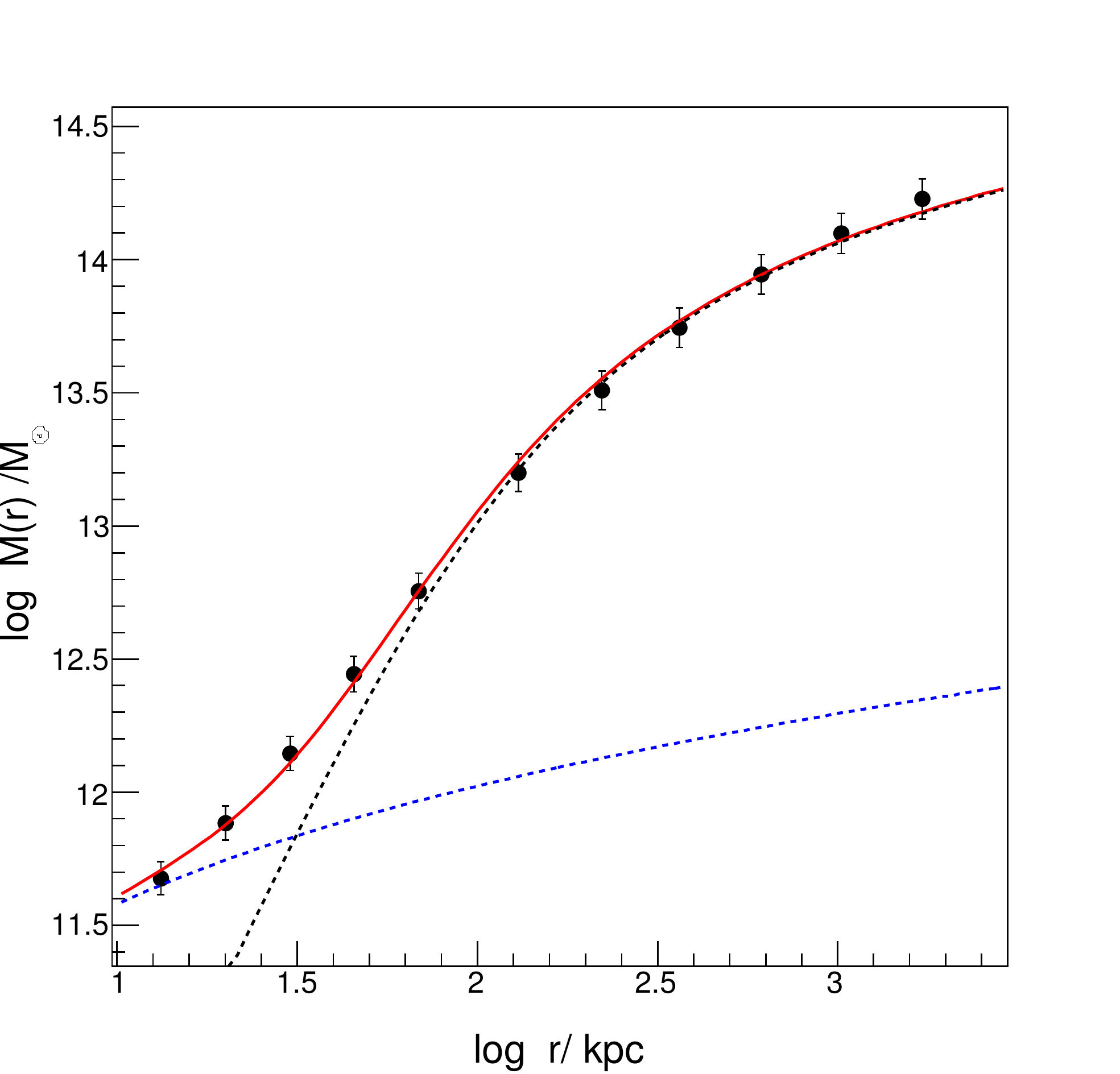}
\includegraphics[scale=0.4]{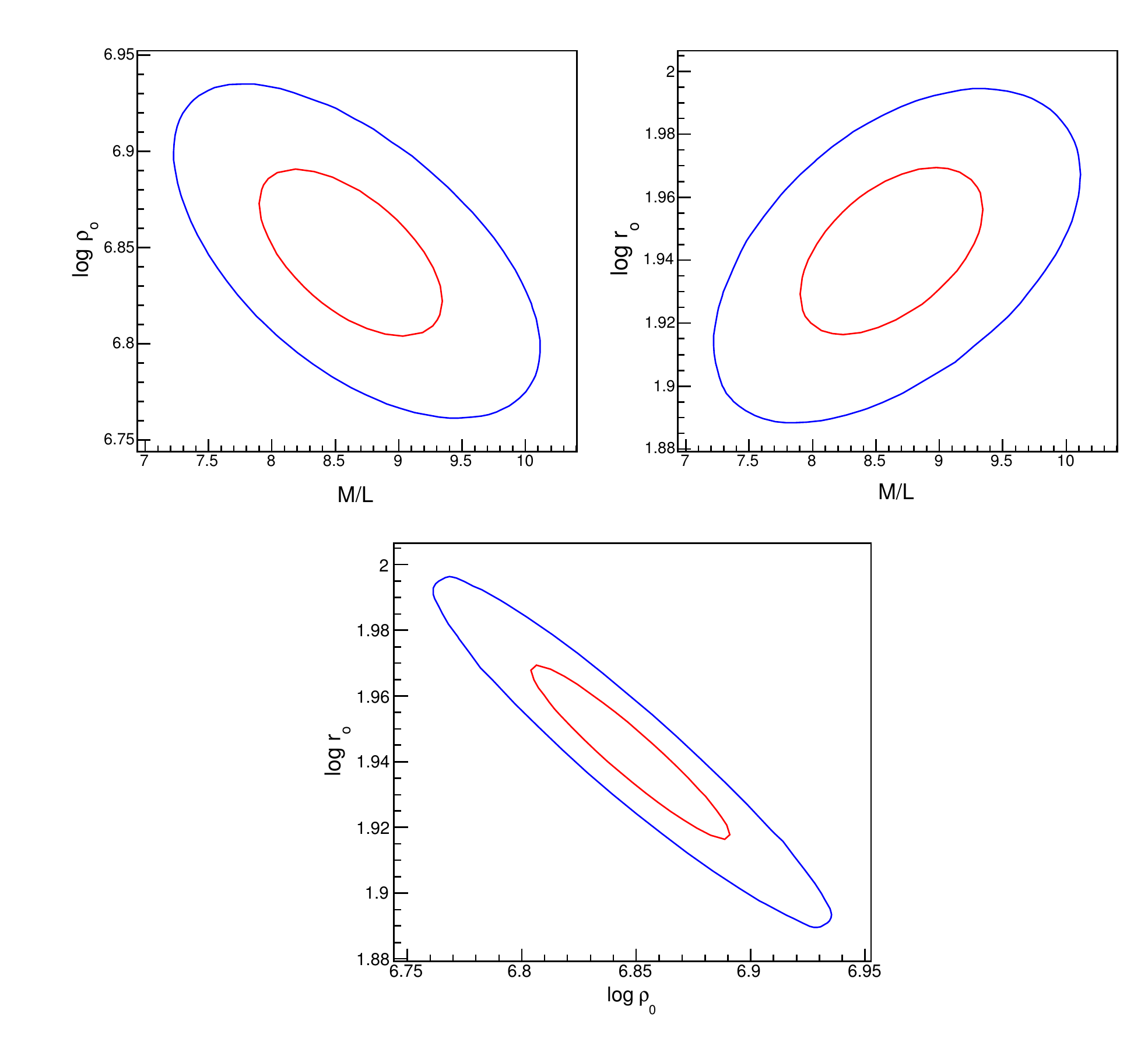}
\caption{The model mass distribution in M87 (red line) also indicated: the  stellar Nuker (dashed blue line) and the Burkert profiles (dashed black line). {\it Bottom}: The 1-sigma (red line) and 2-sigma (blue line) contour plots of the mass model free parameters.}
\label{fig:NukerBukert}
\end{figure}

For the DM, we adopt the Burkert profile; this choice is the only difference concerning the modelling in \cite{Oldham2}, but it is very substantial and, in addition, can also affect the resulting value of the mass of the luminous spheroid. 
Then:
\begin{equation}
\label{rhoh}
\rho_h(r) = \dfrac{\rho_0 r_0^3}{(r+r_0)(r^2+r_0^2)}\,,
\end{equation}
where the core radius $r_0$ and the central density $\rho_0$ are the free parameters of the model. As result: 
\begin{eqnarray}
M_h(r) =&& M_0 \left[\log\left( 1+\frac{r}{r_0}\right)  - \tan^{-1}\left( \frac{r}{r_0} \right)\right.\nonumber\\&&\left. +\frac{1}{2}\log\left( 1+\left( \frac{r}{r_0}\right)^2 \right)\right]\,,
\label{Mass Burkert}
\end{eqnarray}
where $M_0 = 6.4\rho_0 \ r_0^3$. Let us notice that, in the {\it global mass modeling} of such giant galaxy of about $10^{14}M_{\odot}$, the EHT value for the central black hole mass ($(6.5\pm 0.2_{\rm stat}\pm 0.7_{\rm sys} )\times 10^9 M_{\odot}$) shows that the latter does not play any gravitational role (different is the situation in the innermost kpcs, see later).
Therefore, the mass model, with its three free parameters, reads as:
\begin{equation}
\label{Mass Equations} 
M_{mod}(r;M_{sph},r_0,\rho_0) = M_{\star}(r;M_{sph}) + M_h(r; r_0, \rho_0)\,.
\end{equation}
Then, we proceed using the standard $\chi^2$ fitting the eleven independent data of the mass profile $M(r)$ with the mass model of Eq.\eqref{Mass Equations} and obtaining so the three free parameters and the related triangular plots yielding their statistical uncertainties.
It is evident in Fig.\ref{fig:NukerBukert} that the adopted DM profile allied with a much smaller contribution from the stellar spheroid, well reproduces the distribution of the gravitating  mass of the galaxy. Therefore, the mass model can be written as: $M_{mod} (r;M_{sph},\rho_0,r_0)|_{best\,fit}$. 
The resulting values of the best-fit parameters are: $ M_{sph} =(1.3 \pm 0.1)\times 10^{12} \ M_\odot$ that leads to  a mass-to-light ratio of   $M_{sph}/L_V = (8.6 \pm 1.2) M_\odot L_\odot^{-1}$, $r_0 =(91.2 \pm 9.0)$ kpc and $\rho_0=( 4.7 \pm 0.9)\times 10^{-25}$ g/cm$^3$.
The nominal value for the  virial radius  is  $R_{200}=(1.3\pm 0.2)$ Mpc; the  halo mass within $R_{200}$ is $M_{200}^{M87}=(1.3\pm 0.3)\times 10^{14} M_{\odot}$ and the  luminous-dark mass ratio is about $10^{-2}$.  Therefore, M87 is one of the biggest galaxies of the Universe at the upper end of the galaxy formation process. Noticeably, a particle situated at  $R_{200}$, in rotation around the center of M87, would make a complete orbit in not less than $13$\ Gyr, the current age of the Universe, a fact that might be not a coincidence.

\begin{figure}[h!]
\centering
\includegraphics[scale=0.41]{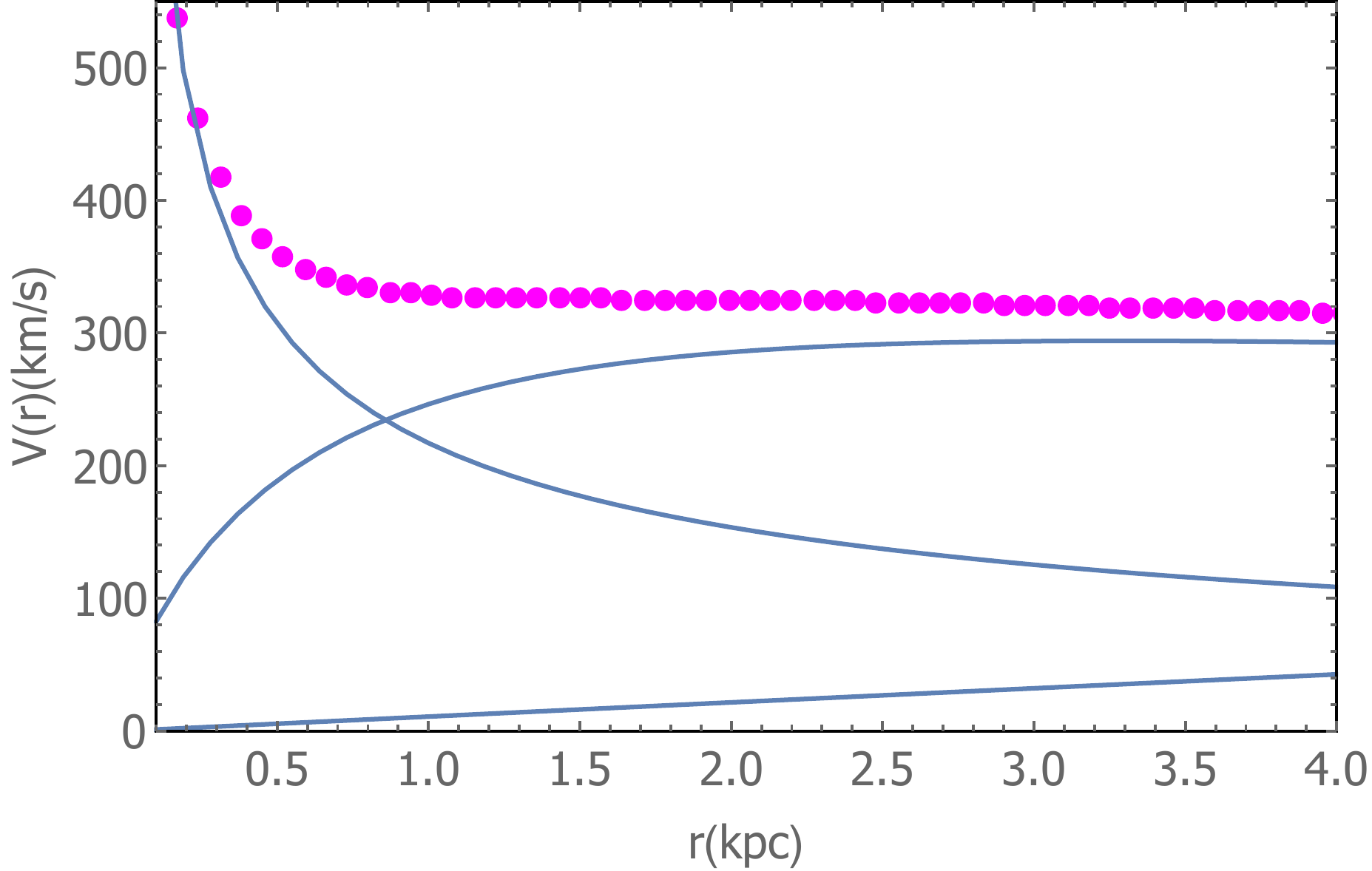}
\caption{The analogous rotation curve in the innermost regions of M87 (violet filled circle). The decreasing and rising curves are the SMBH and spheroid components. Notice: its flat behaviour and the irrelevancy of the dark matter component (the lowest blue line).}
\label{fig:RCM87}
\end{figure}

Let us define, for this spherical pressure dominated object, a circular  velocity analog to that of the rotationally dominated disk systems for which: $V_{M87,disk}^2(r)\equiv G\  M_{M87,disk}(r)/r$
where $V_{M87,disk}(r)$ is the circular velocity that allows a point mass, at a distance $r$ from the center of any galaxy, to stay in rotational equilibrium. 
Then, for M87, considering also Eqs.\eqref{lum mass},\eqref{Mass Burkert}, we have: 
\begin{eqnarray}
G M(r)/r &&= V_{M87}^2(r)= G M_{mod}/r = V_{mod}^2(r)\nonumber\\&&= [V^2_{SMBH}(r)+ V^2_\star(r) +V^2_{h}(r)]_{mod}
\end{eqnarray}
since the model fits the data very well, e.g.: $M_{mod}(r)\simeq M(r)$, we can accurately determine the velocity profile (see Fig.(\ref{fig:RCM87}). Moreover, 
$V(R_{opt})= ( \frac{G M(R_{opt})} {R_{opt}})^{1/2}=360\pm 5\  km/s $ and  for a spherical and  isotropic distribution we derive the dispersion velocity $\sigma(r_e)$ from:  $G^{-1}\, V^2(r_e) r_e =G^{-1}  3 \sigma(r_e)^2\  r_e $ that yields: $\sigma(r_e)= 358\pm 5$ km/s.

Not unexpectedly, the DM density in  the outer parts of M87  ($r>200\,{\rm kpc}$ ) is also well represented by the collisionless NFW profile with a mass of $M^{M87}=1.3 \times 10^{14} \ M_\odot$ and a concentration value of $c=7$. 
This allows us to infer the M87 original DM profile by extrapolating $M^{M87}_{NFW}(r)$ down to $r=0$ so that, defining  $X\equiv r/R_{\rm vir}$, we have: 
\begin{equation}
V_ {NFW}(X)^2=G \ M_{200}/R_{vir} \frac{1}{X} \frac{{{\ln}} (1 + c X) - \frac{c X}{1 + c X}} {{{\ln}} (1 + c) -\frac{c}{1+c}}\,,
\end{equation}
with $M_{200}=200~ 4/3 \pi \ \rho_c \ R_{vir}^3$ and $ \rho_c=1.0 \times 10^{-29}\ {g/cm^3}$. Then , $M_{NFW}(r)=G^{-1}V^2(X R_{vir}) r$

The primordial distribution of the M87 DM halo leads us to realize that a significant fraction of the dark mass, once inside the radius $R_{dom}$\footnote{$R_{dom}$ defines the region in which today the SMBH dominates the dynamics of M87, i.e. $R_{dom}$ = $1$ kpc (see Fig. \ref{fig:RCM87}).}, has gone missing:  $\simeq \Delta M^{M87}_{NFW}(R_{dom})=M_{NFW}^{M87}(R_{dom})-M(R_{dom}) \sim 2 \times 10^9 \ M_\odot$.

A very relevant feature of the mass distribution of M87 is its huge DM core radius $r_0$: this is one of the first detections of DM cores as large as $\sim \ 100$ \ kpc (see also \cite{DiPaolo2019}). Such bare fact was also found by \cite{Oldham2}, although their result could be affected by the various biases of the density profiles adopted.

\begin{figure}
\centering
\includegraphics[scale=0.27]{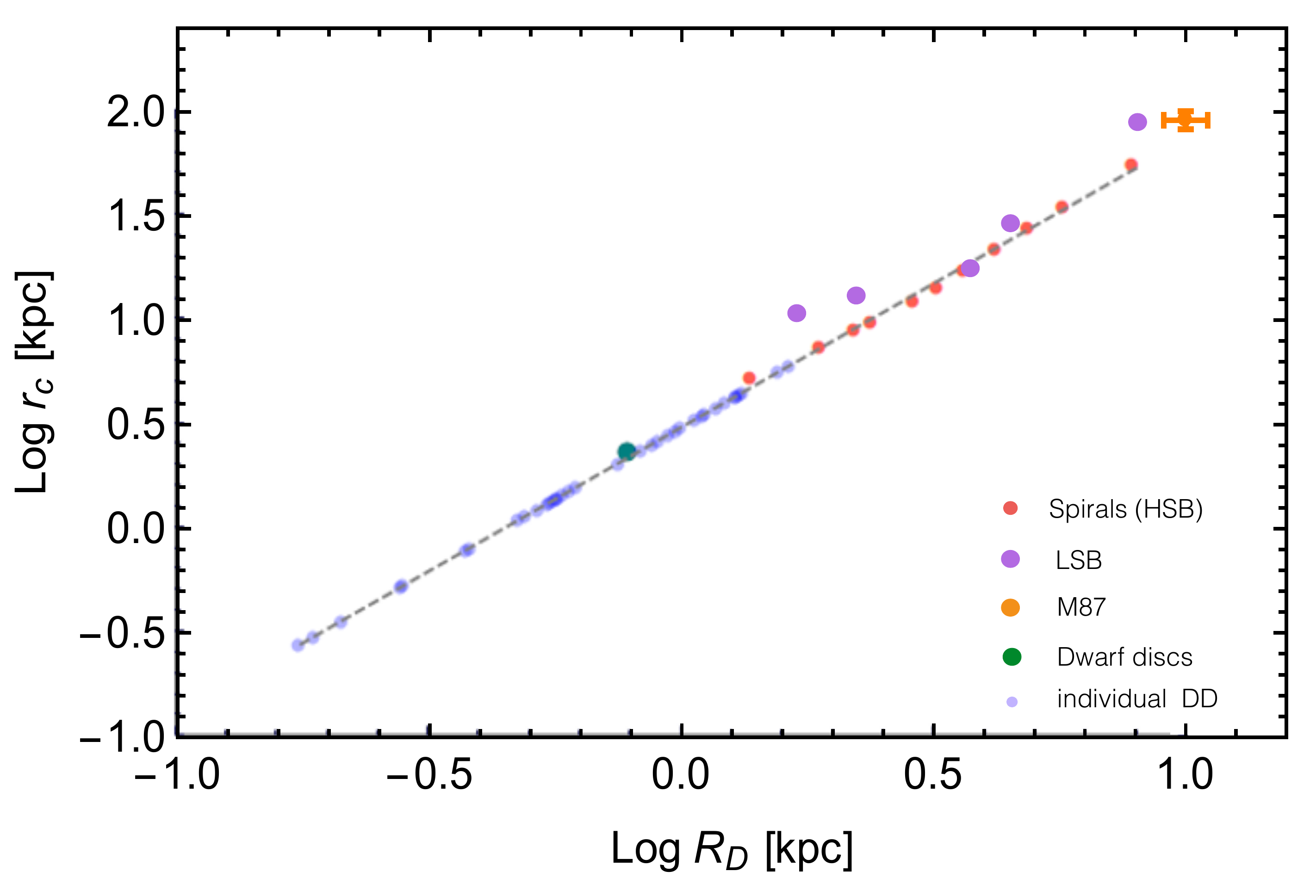}
\caption{The core-radius vs disk  scale length relationship in galaxies (see \cite{Salucci_2019} and references therein).  M87 is represented  given by the orange filled circle. It is  noteworthy the evidence  of core radii ranging from 250 pc to 90 kpc.}
\label{fig:RD}
\end{figure}

M87 has its say also in the  most intriguing relationship of disk galaxies, i.e., that  between the DM core radius $r_0$ and the disk length-scale \citep{Donato:2004af} $R_D$: we realize \citep{Salucci_2019} that it also holds in this  giant galaxy (see Fig. \ref{fig:RD}). In detail, we have that:
$$R_D\equiv\frac{d\,log \Sigma_{\star}}{d\,log\, r }$$
(see Eq. \eqref{Sigma}), a quantity which is deeply connected with the luminous world, strongly correlates with: 
$$r_0\equiv\frac{d\,log \rho_{h}}{d\, log\, r }|_{r=0}$$
a quantity which is, instead, deeply connected with the dark world. Remarkably the connection is at the level of the derivatives of corresponding densities.

\cite{donato2009constant} discovered that, in all spirals, the central DM surface density $\Sigma_0 \equiv \rho_0 r_0$ is about constant:  $\Sigma_0\sim 120^{+200}_{-70}\  M_\odot/$pc$^2$, independently of the galaxy magnitude. 
Such relationship has been later confirmed across a range of $18$ magnitudes (maximum circular velocity) and in galaxies of different Hubble types: dwarf spheroidal, dwarf irregulars and LSBs \citep{salucci2012dwarf} (see also \cite{Salucci_2019}).   
We realise that M87 follows such a dark world  - dark world quantities relationship within a factor 4. 
Furthermore, M87, with a stellar mass much higher than any other galaxy,  is pivotal in indicating that in the above relationship also the stellar mass plays a role. A more accurate relationship involving two quantities of the DM component, $\rho_0$ and $r_0$ and one quantity of the luminous matter component $M_{sph}$, emerges
 ($\Sigma_0$ in $M_\odot/pc^2$):
\begin{equation}
\log \ \Sigma_0= 2.72 + \frac{1}{6} \log \Big(\frac {M_{sph}}{10^{12} \ M_\odot}\Big)\,, 
\end{equation}
with the scatter less of 0.15 dex.

\section{The central super massive black hole as a mass component of M87}
\label{3}
The accurate mass model of M87 of previous Sec. \ref{2} and the very precise EHT determination of the mass of its SMBH \citep{Akiyama:2019cqa,Akiyama:2019eap} allow us to investigate the region in which the latter dominates the galaxy gravitational potential. 
First, we notice that, compared with other spheroidals, the  fraction between the SMBH mass and  the total stellar mass is quite  high:  $6. \times 10^{-3}$, especially considering that most of the $1.2\times 10^{12} M_\odot$ of the M87 stellar spheroid has been accumulated at times much later than that of the formation of the SMBH. 
Using the full M87 mass model, we can derive the analogous circular velocity $V(r)$  from  $r\sim 0.1$\ kpc outwards, determining the various luminous, black hole and dark matter components. 
Let us stress that, in the region $ 0.1$\ kpc $<r< 10$ \ kpc,  we recover $V(r)$  without having in this region dynamical measurements. The DM component here is negligible. The EHT independently measured the mass of the SMBH. Finally, this region's stellar mass distribution is obtained through high-resolution photometry alongside dynamical measurements at 10-30 Kpc, where the stellar spheroids are the major massive component.

In Fig. \ref{fig:RCM87} we plot such curve out to $4$ kpc; we realize that inside 1 kpc, the SMBH component dominates that of the stellar halo and, therefore, the whole gravitational potential. Remarkably, such component gives an important contribution to the total velocity out to $4$\  kpc $\simeq 0.4 \ R_D$. 
This result is amazing: in late Spirals, of any mass, the radius of the dominance of the SMBH is lesser than $20-50$ \ pc  $\sim 0.05 \ R_D$ \citep{Salucci:1998ij}. 
In Elliptical galaxies, we do not see a dynamically dominant SMBH component in that the SMBHs with masses of $10^{8-9}\  M_\odot$ are buried inside stellar spheroids  with a mass of $>10^{11}\ M_\odot$ within $r_e$\footnote{Of course, the accretion-disk related maser sources are totally subject to the SMBH gravitational field.}. 
To our knowledge, M87 is the unique object in which we can see the central black hole participating, with another mass component, in shaping the mass distribution of a galaxy. 
The two components, actually, looks fine-tuned and the circular velocity of M87 keeps constant from 0.4 kpc to 4 kpc in spite of the fact that they, in such region, have totally different velocity radial profiles: $V_{SMBH}(r) \propto r^{-1/2}$ which contrasts $V_{\star}(r) \propto r^{1/2}$.

Another peculiar feature seems present in this region: a mysterious lack of DM. 
In fact, inside $0.1\  R_D=1$\ kpc the extrapolated back dark mass $M_{NFW}^{M87}(1 \ {\rm kpc})$ is only  $0.1\% $ of  $M(1\ {\rm kpc})$, computed  from the M87* black hole mass and from accurate photometry and the mass to light ratio of the stellar spheroid. Inside $R_D$, where we have dynamical measurements of the total mass, the dark mass is  $13\%$ of $M(R_D)$. Only at $r> 22$\ kpc  the dark matter contribution to the circular velocity overcomes that of the standard matter.

In spheroidals the well known $M_{BH}$ vs  $log \ \sigma_e $ relationship is a fundamental one e.g. \cite{Kormendy:2013dxa}):
\begin{equation}
\label{8}
\log \frac {M_{BH}}{10^9\  M_\odot} =-0.51\pm 0.05 + (4.4 \pm 0.03) \log \frac{\sigma_e}{200 \ km/s}\,.
\end{equation}

Since in M87 $M_{BH}=6.5 \times 10^9 \ M_\odot$ and $\sigma_e \simeq  (358 \pm 5) {\rm km/s} $, the EHT SMBH mass results $(0.2 \pm 0.04)$ dex larger than the value predicted by Eq. \eqref{8}, showing a $5 \sigma$ excess. Such excess, in mass (at $1\sigma$) is of the order of $(3-4.5) \times 10^9 M_\odot$. This is an important result, given the exquisite EHT black hole mass estimate and the fact that the \cite{Kormendy:2013dxa} relationship is very tight.  In fact, in the crucial process of the black hole mass growth, this result points to a role for the dark matter. This may be the first case in which one can argue that the DM component also contributed to the SMBH mass growth.

\section{Discussion and Conclusions}
\label{4}
For several issues of Cosmology and extra-galactic astrophysics the determination of the M87 dark and standard matter mass distributions of this work becomes a phenomenal test-bed and yields pivotal information.  
M87 is a giant cD galaxy of the total mass of $ 1.3 \times 10^{14} \ M_{\odot}$ which has, at its center, the usual SMBH, whose mass has been exquisitely measured in the process of imaging its shadow.  
In addition, we have an accurate mass distribution from 100\ pc to 1 Mpc obtained by exploiting high quality photometric and spectroscopic measurements. 

Remarkably, the M87* SMBH, almost uniquely in the local Universe,  controls gravitationally a region $\simeq 1 $ kpc wide, populated by more than $10^7$ stars of the stellar spheroid. 
In contrast, in our Galaxy, the central SMBH hole SgrA* dictates the motion of only a few thousand stars. 

There is evidence that, inside $R_{dom} \simeq 1$ kpc, a relevant portion  of the original DM halo mass has disappeared over the Hubble Time, in fact: $M_{DM} (R_{dom},z_{form} )-M_{NFW}(R_{dom}, 0) \sim 2\times  10^9 M_\odot$.  
On the other side, the EHT estimate of the M87* mass results bigger, by approximately the same amount, than the value expected from the well known $M_{SMBH}$ vs $Log \ \sigma_e$ diagram, that people think born out of an Eddington accretion of standard matter onto a SMBH seed. 
The following argument also supports the presence of a dark infall on M87*. In the case of a Dark halo around a SMBH of $\sim 7 \times 10^9 M_\odot$ whose density at $r=R_{dom} $ is  $\rho_{DM}{R_{dom}}$ as found in the previous section, develops, for $r<R_{dom}$ a cusp so that: $\rho_{DMcusp}(r)= \rho_{DM}{R_{dom}} (r/R_{DM})^{-2.5}$
The mass inside this hypothetical cusp will be $2 \times 10^{11}$ totally incompatible with the DM mass dynamically estimated at 20 kpc.

These features may lead to the first observed case in which one can argue that the DM component also contributed to the SMBH mass growth. More generally, one can envisage that inside the innermost hundreds of parsec, the escape velocity from the giant black hole is much bigger than the dispersion velocity of the primordial Dark Matter halo particles. On this line, it is known that the capture of DM particles by the central black hole is a well studied physical process see e.g. \cite{Gammaldi:2016uhg}. Furthermore, let us remind that the energy and angular momentum of the DM particles in the innermost kpc can be removed by the non-collisional status of the particles itself: e.g. in the scenario of self-interacting dark matter particles and in that of interacting dark matter-standard model particles e.g.  \cite{Salucci:2020eqo}.

The SMBH seems to have been, over the cosmological times, dynamically "live" in a surprising way: a "fine-tuning" between the mass of M87* SMBH and the mass {\it distribution} of its stellar spheroid appears. In fact, in the region inside 4 \ kpc the effective circular velocity keeps constant, despite that the stellar spheroid and the SMBH components have very different radial profiles and total masses. 
We can argue that the central SMBH has dynamically shaped the innermost portion of the stellar spheroid.

The $1 \times 10^{14} \ M_\odot$ massive M87 DM halo density is well reproduced by a Burkert profile with a $\sim 100$ \ kpc wide core radius. 
Outside the region in which the dark halo coexists with the stellar spheroid, $r> 200$ kpc, the DM density converges to the NFW profile characteristic of the collisionless  DM regime, as it occurs in Spirals \citep{Salucci_2019}. 

M87 is a benchmark for the idea that supernovae (SN) induced baryonic feedbacks, by flattening the original cuspy distribution, are the cause of the detected DM cores. 
The energy budget of this possible process can be easily computed.
The (potential) energy $E_{cf}$ involved in the core forming process, taking into account that, inside $r_0$, $\rho_{NFW}^{M87}(r)>>\rho_{DM}^{M87}(r)$ is: $E_{cf}\sim 10^{62}$ \ erg. 
The stellar mass inside $r_0$ is about $\sim 8 \times 10^{11} M_\odot $ that implies, for a single burst of star formation with a Salpeter IMF, the explosions of $\sim 5 \times 10^9$ core collapse SNs, each of them providing an energy of $ f_{cf} 10^{51} $ erg to the core forming process, where $f_{cf}\sim  10^{-1} - 10^{-2} $. 
The cumulative feedback energy, therefore, does not reach $10^{60}$\ erg $<< E_{cf}$. 
Notice that also in the biggest spirals the SN feedback is short of providing sufficient energy, but in M87 the failure is outstanding.

The stellar spheroid and the DM halo main structural length scales, $R^{M87}_D$ and $r_0$, already mysteriously related in disk systems of any mass and morphology \citep{Salucci_2019,Salucci:2020eqo}, continue to be clone-like entangled also in this extremely different galaxy.

The M87 DM structural properties result decisive in one important test about the nature of the dark particles. A well preferred solution for the riddle of the observed cored DM distributions around galaxies involves the role of the  Fuzzy Dark Matter (FDM) , a hypothetical particle (e.g. an Ultra Light Axion)  with a mass of the order of $m_p =10^{-22}$\ eV that implies a de Broglie wavelength on the galaxy scales. FDM halos well reproduce  the kinematics of Dwarf galaxies with halo masses  $ M_h \sim 10^{9-10} M _\odot $ e.g.: \cite{Schive:2014hza,Hui:2016lt,deMartino:2020gfi,Pozo2020} and references therein. At larger masses the situation is completely open. However, in combination with the recent result by \cite{Burkert2020} according to which $\Sigma_0$, $r_0$ and the galaxy redshifts of halo formation $z_{form}$ are related by: 
\begin{equation}
\label{9}
    \frac{r_0^3}{\rm kpc^3}=0.25 (1+z_{form})\frac{75\ M_\odot pc^{-2}}{\Sigma_0} \frac{10^{-22} eV}{m_p}
\end{equation}
the present result draw light on an important feature. 
For our galaxy, of mass $M_{200}^{M87}=1.3 \times 10^{14} \ M_\odot$, we have: $z_{form}<5$, $\Sigma_0 \sim \ 500 \ M_\odot$ pc$^{-2}$ and $r_0\sim 90 \ kpc$, then Eq. \eqref{9} holds only for particle masses whose de Broglie wavelength are at the level of a Mpc scale and are so unable to account for the DM halo density cores.  
Considering the family of normal spirals, 
\cite{Burkert2020} noticed the implausibility, in the current galaxy formation theory, of the $r_0$ $\propto$  $(1+z_{form})^{1/3}$ relationship in Eq. \eqref{9}. Overall, in M87 we conclude that the inferred value of $r_0$ results totally inconsistent with the prediction of Eq. \eqref{9}: the latter cannot be claimed to dictate the size of the constant density region in galaxies.

M87 provides evidence that the primordial DM halo distribution has been modified by the combined action of the central SMBH and the stellar spheroid during the entire life of the Universe.  Therefore, it could be the place in which a new paradigm for the dark matter phenomenon arises: the latter's nature will emerge by reverse engineering the entanglement among the dark-luminous structural properties that we detect in galaxies rather than coming from theoretical first principles.


\bibliography{main.bbl}
\bibliographystyle{aasjournal}



\end{document}